\documentclass[12pt]{article}
\usepackage{epsf}
\usepackage[dvips]{graphicx,psfrag}

\renewcommand{\thefootnote}{\#\arabic{footnote}}
\renewcommand{\theequation}{\thesection.\arabic{equation}}

\begin{document}

\newcommand{\gtrsim}{ \mathop{}_{\textstyle \sim}^{\textstyle >} }
\newcommand{\lesssim}{ \mathop{}_{\textstyle \sim}^{\textstyle <} }

\renewcommand{\thefootnote}{\fnsymbol{footnote}}
\setcounter{footnote}{0}
\begin{titlepage}

\def\thefootnote{\fnsymbol{footnote}}

\begin{center}

\hfill TU-717\\
\hfill hep-ph/0404253\\
\hfill April, 2004\\

\vskip .5in

{\Large \bf

Curvaton Scenario with Affleck-Dine Baryogenesis

}

\vskip .45in

{\large
Maki Ikegami and Takeo Moroi
}

\vskip .45in

{\em
Department of Physics, Tohoku University, Sendai 980-8578, Japan
}

\end{center}

\vskip .4in

\begin{abstract}

We discuss the curvaton scenario with the Affleck-Dine baryogenesis.
In this scenario, non-vanishing baryonic entropy fluctuation may be
generated even without primordial fluctuation of the Affleck-Dine
field.  Too large entropy fluctuation is inconsistent with the
observations and hence constraints on the curvaton scenario with the
Affleck-Dine baryogenesis are obtained.  We calculate the baryonic
entropy fluctuation (as well as other cosmological density
fluctuations) in this case and derive constraints.  Implications to
some of the models of the curvaton are also discussed.

\end{abstract}

\end{titlepage}

\renewcommand{\thepage}{\arabic{page}}
\setcounter{page}{1}
\renewcommand{\thefootnote}{\#\arabic{footnote}}
\setcounter{footnote}{0}

\renewcommand{\theequation}{\thesection.\arabic{equation}}

\section{Introduction}
\label{sec:introduction}
\setcounter{equation}{0}

Recent precise measurement of the cosmic microwave background (CMB)
anisotropies by the Wilkinson Microwave Anisotropy Probe (WMAP)
\cite{wmap} has provided detailed informations about the mechanism of
generating the cosmic density fluctuations.  Importantly, observed CMB
anisotropies are highly consistent with those predicted from
scale-invariant purely adiabatic primordial density fluctuations.
Such a class of primordial density fluctuations is also consistent
with the recent results by the Sloan Digital Sky Survey experiments
\cite{Tegmark:2003ud}.  Among various possibilities, inflation is one
of the most famous scenarios of generating cosmic density fluctuations
consistent with the observations.

Another class of scenario of generating the scale-invariant adiabatic
density fluctuations, however, exists, which is called the
``curvaton'' scenario
\cite{Enqvist:2001zp,Lyth:2001nq,Moroi:2001ct}.\footnote
{A similar study of the effect of extra scalar field other than the
inflaton, see also \cite{PRD56-535}.}
(For the recent works on the curvaton scenario, see
\cite{RecentCurvaton}.)  If there exists a scalar field $\phi$ (other
than the inflaton) which acquires primordial fluctuation during the
inflation, dominant part of the cosmological density fluctuations may
originate from the primordial fluctuation of this scalar field.  (The
scalar field $\phi$ is called the curvaton.)  Importantly, this
scenario works irrespective of the detailed properties of the curvaton
field, i.e., its mass, lifetime, and initial amplitude.  Consequently,
there are many possible and well-motivated candidates of the curvaton
field since, in various scenarios of particle cosmology, many
scalar-field condensations which once dominate the universe show up;
if those scalars acquire primordial quantum fluctuations, they may
play the role of the curvaton.  Thus, the curvaton scenario naturally
fits into various scenarios of particle cosmology.  In addition, with
the curvaton scenario, we have a chance to test the properties of the
particle responsible for the cosmological density fluctuations by
collider experiments; this is the case if the flat direction of a
minimal supersymmetric standard model (MSSM) becomes the curvaton, for
example.  (Remember that it is difficult to construct a model of
inflation within the framework of the standard model or the
MSSM.\footnote
{In the MSSM, however, it may be possible to use the $D$-flat
direction consisting of the up-type squarks as well as the up-type
Higgs boson as the inflaton.  For details, see \cite{Kasuya:2003iv}.})

An important aspect of the curvaton scenario is that the universe is
reheated by the decay of the curvaton.  In some case, reheating
temperature at the curvaton decay becomes relatively low and hence it
is non-trivial if a viable scenario of baryogenesis can be found.  In
addition, a large amount of entropy is produced at the time of the
decay of the curvaton.  As a result, even if non-vanishing baryon
asymmetry is generated before the decay of the curvaton, primordial
baryon asymmetry may be too much diluted to be consistent with the
present value.  This may be a problem for the case where, for example,
the cosmological moduli fields play the role of the curvaton.

One of the possibilities to generate enough baryon asymmetry with
large entropy production is to adopt the Affleck-Dine baryogenesis
\cite{Affleck:1984fy}.  Indeed, in \cite{Moroi:1994rs}, it was pointed
out that the resultant baryon-to-photon ratio $n_b/n_\gamma$ (with
$n_b$ and $n_\gamma$ being the number densities of the baryon and
photon, respectively) can be as large as the presently observed value
$\sim O(10^{-10})$ even with the large late-time entropy production.
In the Affleck-Dine scenario, however, baryonic isocurvature
fluctuation may arise since, in this scenario, baryon asymmetry is
generated from a coherent motion of a scalar field (called
Affleck-Dine field).  In particular, if the Affleck-Dine mechanism is
implemented in the curvaton scenario, correlated baryonic isocurvature
fluctuation may arise as we will see below.  Thus, it is important to
study constraints on this case and see if the curvaton scenario with
Affleck-Dine baryogenesis is viable.

In this paper, we consider the curvaton scenario with the Affleck-Dine
baryogenesis.  In particular we study the possible baryonic entropy
fluctuation generated in this framework and how the entropy
fluctuation can be suppressed.  The organization of the rest of this
paper is as follows.  In Section \ref{sec:evolutions}, we first
discuss framework and give the thermal history we consider.  Then, we
introduce equations by which evolutions of the cosmological density
fluctuations are governed.  Analytic solutions to those equations
for a simple (but well-motivated) case are also discussed.  In Section
\ref{sec:numerical}, we numerically solve the evolution equations.  In
particular, we will see that the baryonic entropy fluctuation provides
very stringent constraints on the curvaton scenarios with Affleck-Dine
baryogenesis.  Implications of such constraints for some of the
curvaton scenarios are discussed in \ref{sec:implications}.  Section
\ref{sec:summary} is devoted to the summary of this paper.

\section{Evolutions of the Fluctuations}
\label{sec:evolutions}
\setcounter{equation}{0}

\subsection{Thermal history}

Let us first introduce the framework.  In various models of physics
beyond the standard model, there are many scalar fields whose
potential is approximately parabolic:
\begin{eqnarray}
    V(\phi) = \frac{1}{2} m_\phi^2 \phi^2.
\end{eqnarray}
In the MSSM, for example, $F$- and $D$-flat directions, whose
potential is lifted by the supersymmetry-breaking effects, has
parabolic potential.  In addition, potential of the cosmological
moduli fields are expected to be parabolic when the amplitude of the
modulus field is smaller than $\sim M_*$ (where $M_*$ is the reduced
Planck scale).  Another possible candidates of such a scalar field is
pseudo-Nambu-Goldstone bosons \cite{PNGBcurvaton}.  In this paper, we
assume one of such scalar fields plays the role of the curvaton.

Here, we adopt inflation as a solution to the horizon, flatness, and
other cosmological problems.  Then, the universe starts with the
inflationary epoch where the universe is dominated by the potential
energy of the inflaton field $\chi$; thermal history before the
inflation is irrelevant for our discussion.  During the inflation, we
assume that the amplitude of the curvaton is non-vanishing.  In
addition, we consider the case where the effective mass of the
curvaton during the inflation is much smaller than the expansion rate
during the inflation $H_{\rm inf}$.  Then, the curvaton acquires the
primordial fluctuation during the inflation
\begin{eqnarray}
    \delta \phi_{\rm init} = \frac{H_{\rm inf}}{2\pi}.
    \label{dphi_init}
\end{eqnarray}
(Here and hereafter, the subscript ``init'' is used for initial values
of the quantities at very early epoch when $\phi$ is slowly rolling.)
After the inflation, universe is reheated by the decay of the inflaton
field and the universe is dominated by radiation.  We call this epoch
the first radiation-dominated epoch (or RD1 epoch) since, in our
scenario, there are two radiation-dominated epochs.

As far as the decay of $\phi$ is negligible, universe consists of two
components: radiation generated from the decay products of the
inflaton, which we denote $\gamma_\chi$, and the curvaton.  After the
RD1 epoch, evolution of the energy density of $\gamma_\chi$ obeys the
following equation:
\begin{eqnarray}
    \dot{\rho}_{\gamma_\chi} + 4 H \rho_{\gamma_\chi} = 0,
    \label{dot(rho_r)}
\end{eqnarray}
where the ``dot'' denotes the derivative with respect to the time $t$.
In addition, (unperturbed part of) $\phi$ obeys the following
equation:
\begin{eqnarray}
    \ddot{\phi} + 3 H \dot{\phi} + m_\phi^2 \phi = 0.
\end{eqnarray}
(As is obvious from the above equations, we have not taken into
account of the decay of the curvaton field since the effect of the
decay of $\phi$ is negligible for the RD1 and $\phi$-dominated
epochs.)  Here, $H\equiv \dot{a}/a$ (with $a$ being the scale factor)
is the expansion rate of the universe which is given by, when 
$H$ is much larger than the decay rate of $\phi$,
\begin{eqnarray}
    H^2 = \frac{1}{3M_*^2} 
    \left( \rho_{\gamma_\chi} + \frac{1}{2}\dot{\phi}^2 + 
        \frac{1}{2} m_\phi^2 \phi^2 \right).
\end{eqnarray}

From Eq.\ (\ref{dot(rho_r)}), we can see that
$\rho_{\gamma_\chi}\propto a^{-4}$.  On the contrary, the curvaton
starts to oscillate when $H\sim m_\phi$; after this epoch, energy
density of the curvaton decreases as $\rho_\phi\propto a^{-3}$.  Thus,
the universe can be dominated by the curvaton if the lifetime of the
curvaton is long enough.  Hereafter, the $\phi$-dominated epoch is
called as $\phi$D epoch.  After the $\phi$D epoch, the curvaton decays
and the universe is reheated again; we call the second
radiation-dominated epoch as RD2 epoch.  After the RD2 epoch, the
thermal history is the same as that of the conventional big-bang model
(as far as the zero-mode is concerned).

The curvaton amplitude $\phi_{\rm eq}$ at the time of the
``radiation-curvaton equality'' is easily estimated.  (In the
following, the subscript ``eq'' is used for quantities at the time of
the radiation-curvaton equality.)  The curvaton starts to oscillate
when the temperature of the radiation generated by the inflaton decay
(which is denoted as $\gamma_\chi$ hereafter) becomes
$\sim\sqrt{m_\phi M_*}$ and, after this epoch, the curvaton amplitude
is approximately proportional to $a^{-3/2}$.  Thus,
\begin{eqnarray}
    \phi_{\rm eq} = \phi_{\rm init}^4 M_*^{-3}.
\end{eqnarray}
Expansion rate at the equality is also easily estimated:
\begin{eqnarray}
    H_{\rm eq} \sim 
    \frac{m_\phi\phi_{\rm eq}}{M_*}
    \sim
    m_\phi \left( \frac{\phi_{\rm init}}{M_*} \right)^4.
    \label{H_eq}
\end{eqnarray}

Assuming that the energy density of the Affleck-Dine field is always
sub-dominant in the RD1 and $\phi$D epochs, which we assume in this
paper, the above discussion holds even with the Affleck-Dine field.
Equation of motion of the Affleck-Dine field is, on the other hand,
given by\footnote
{In our convention, the Affleck-Dine field $\psi$ is a complex scalar
field while the curvaton $\phi$ is a real scalar field.  We assume
that the effect of the motion of the curvaton in the phase direction
is negligible.  (For the effect of the motion of the curvaton in the
phase direction, see \cite{HamKawMorTak}.)}
\begin{eqnarray}
    \ddot{\psi} + 3 H \dot{\psi} +
    \frac{\partial V_{\rm AD}}{\partial\psi^*} = 0,
\end{eqnarray}
where $V_{\rm AD}$ is the potential of the Affleck-Dine field.

Evolution of the Affleck-Dine field depends on the detailed shape of
its potential.  In this paper, to be specific, we consider the case
where an $F$- and $D$-flat direction in the MSSM plays the role of the
Affleck-Dine field.  In this case, baryon-number violating interaction
in the potential of $\psi$ is assumed to originate from K\"ahler
interaction.  Another possibility is with baryon- (or lepton-) number
violating superpotential.  Then, however, the initial amplitude of the
Affleck-Dine field is usually suppressed.  Consequently the resultant
baryon asymmetry becomes too small to be consistent with the presently
observed value if a large amount of entropy is produced by the decay
of the curvaton.  One of the motivation here to consider the
Affleck-Dine baryogenesis is to generate large enough baryon-number
asymmetry.  In the case of the $F$- and $D$-flat direction as the
Affleck-Dine field, it is known that baryon asymmetry can be large
enough even with the large entropy production \cite{Moroi:1994rs}.
Thus, we consider the Affleck-Dine potential of the form:
\begin{eqnarray}
    V_{\rm AD} = 
    m_\psi^2 |\psi|^2 + \frac{\lambda m_\psi^2}{M_*^{p-2}} 
    \left( \psi^p + \psi^{*p} \right),
\end{eqnarray}
where $m_\psi^2$ is from the supersymmetry-breaking effect and
$\lambda$ is a constant of $O(1)$.  Then, potential of the
Affleck-Dine field is dominated by the parabolic term as far as
$|\psi|\lesssim M_*$ and $\psi$ starts to oscillate when $H\sim
m_\psi$.  After this epoch, $|\psi|$ is approximately proportional to
$a^{-3/2}$.

Denoting the baryon-number of $\psi$ as $B_\psi$, relevant part of the
baryon-number current is given by
\begin{eqnarray}
    n_b = i B_\psi \left( 
        \psi^* \dot{\psi} - \dot{\psi}^* \psi
    \right).
\end{eqnarray}
Then, we obtain
\begin{eqnarray}
    \dot{n}_b + 3 H n_b = i B_\psi
    \left( \frac{\partial V_{\rm AD}}{\partial\psi} \psi
        - {\rm h.c.} \right).
\end{eqnarray}
Thus, if the Affleck-Dine field has non-vanishing initial amplitude
$\psi_{\rm init}$, baryon-number density is generated when $\psi$
starts to move.  The Affleck-Dine field starts to move when $H\sim
m_\psi$.  Parameterizing the initial value of 
the Affleck-Dine field as
$\psi_{\rm init}\equiv |\psi_{\rm
init}|e^{i\theta_{\rm init}}$, the baryon-number density at that
moment is estimated as
\begin{eqnarray}
    \left[ n_b \right]_{H\sim m_\psi} \sim
    \left[ B_\psi {\rm Im} 
        \left( \frac{\partial V_{\rm AD}}{\partial\psi} \psi 
        \right)
        H^{-1} \right]_{H\sim m_\psi} 
    \sim
    \lambda B_\psi \frac{m_\psi}{M_*^{p-2}} 
    |\psi_{\rm init}|^p \sin  p\theta_{\rm init},
    \label{n_b(H=m)}
\end{eqnarray}
where coefficients of $O(1)$ is neglected.

First, let us consider the case where the present CMB radiation is
generated from the decay product of the curvaton.  In this case, the
resultant baryon-to-photon ratio depends on the epoch when the
slow-roll condition of the Affleck-Dine field breaks down.  (Here, we
consider the case with $m_\psi\lesssim m_\phi$; as we will see below,
this relation is required in order to suppress the baryonic entropy
fluctuation.)  To discuss the resultant baryon-to-photon ratio, it is
convenient to define
\begin{eqnarray}
    \phi_{\rm crit} = M_* \left( \frac{m_\psi}{m_\phi} \right)^{1/4}.
\end{eqnarray}
If $\phi_{\rm init}\ll\phi_{\rm crit}$, $H\sim m_\psi$ is realized in
the RD1 epoch while, for the case of $\phi_{\rm init}\gg\phi_{\rm
crit}$, the Affleck-Dine field starts to oscillate in the $\phi$D
epoch.

If the Affleck-Dine field starts to move in the RD1 epoch, background
temperature at $H\sim m_\psi$ is estimated as $\sim\sqrt{m_\psi M_*}$.
At this moment, with the condition $m_\psi\lesssim m_\phi$, amplitude
of $\phi$ is given by $\phi\sim\phi_{\rm init}(m_\psi/m_\phi)^{3/4}$.
Then, the ratio $n_b/\rho_\phi$, which is a conserved quantity when
$H\lesssim m_\psi$, is estimated as
\begin{eqnarray}
    \frac{n_b}{\rho_\phi} \sim 
    \frac{\lambda B_\psi |\psi_{\rm init}|^p \sin p\theta_{\rm init}}
    {m_\phi^{1/2} m_\psi^{1/2} M_*^{p-2} \phi_{\rm init}^2}.
\end{eqnarray}
When the curvaton decays, energy density of $\phi$ is converted to
that of radiation and hence the baryon-to-photon ratio is estimated as
\begin{eqnarray}
    \frac{n_b}{n_\gamma} \sim 
    \frac{\lambda B_\psi |\psi_{\rm init}|^p \sin p\theta_{\rm init}}
    {m_\phi^{1/2} m_\psi^{1/2} M_*^{p-2} 
    \phi_{\rm init}^2} T_{\rm RD2},
    \label{eta_1}
\end{eqnarray}
where $T_{\rm RD2}$ is the reheating temperature at the time of the
curvaton decay.  Thus, in this case, the resultant baryon-to-photon
ratio depends on the initial amplitude of the curvaton field.  If
$H\sim m_\psi$ is realized in the $\phi$D epoch, on the contrary, the
situation changes and the expansion rate at the time when $\psi$
starts to move is determined by the energy density of the curvaton.
Consequently, at the time of $H\sim m_\psi$, amplitude of the curvaton
is estimated as
\begin{eqnarray}
    [ \phi ]_{H\sim m_\psi} \sim M_* 
    \left( \frac{m_\psi}{m_\phi} \right).
\end{eqnarray}
Then,  the baryon-to-photon ratio is estimated as
\begin{eqnarray}
    \frac{n_b}{n_\gamma} \sim 
    \frac{\lambda B_\psi |\psi_{\rm init}|^p\sin p\theta_{\rm init}}
    {m_\psi M_*^p} 
    T_{\rm RD2}.
    \label{eta_2}
\end{eqnarray}
In this case, the resultant baryon-to-photon ratio does not depend on
$\phi_{\rm init}$.

If the lifetime of the Affleck-Dine field is long enough, the present
CMB radiation dominantly originate from the decay product of the
Affleck-Dine field.  In this case, the expression of the
baryon-to-photon ratio changes using the fact that the ratio
$n_b/\rho_\psi$ (with $\rho_\psi$ being the energy density of the
Affleck-Dine field) is a constant after the Affleck-Dine field starts
to move.  The baryon-to-photon ratio is estimated as
\begin{eqnarray}
    \frac{n_b}{n_\gamma} \sim 
    \left[ \frac{n_b}{\rho_\psi} \right]_{H\sim m_\psi} 
    T_{\rm ADdecay}
    \sim 
    \frac{\lambda B_\psi |\psi_{\rm init}|^{p-2}
    \sin p\theta_{\rm init}}
    {m_\psi M_*^{p-2}} 
    T_{\rm ADdecay},
    \label{eta_3}
\end{eqnarray}
where $T_{\rm ADdecay}$ is the reheating temperature due to the decay
of the Affleck-Dine field.  Then, the baryon-to-photon ratio is
independent of $\phi_{\rm init}$ irrespective of the time when the
Affleck-Dine field starts to move.

\subsection{Adiabatic density fluctuations}

Now, we consider the evolution of the density fluctuations.  In our
scenario, there are two independent sources of the cosmological
density fluctuations; one is the primordial fluctuation of the
curvaton given in Eq.\ (\ref{dphi_init}) and the other is that of the
inflaton.  Since we are interested in the case where the curvaton
contribution dominates, let us consider the evolution of the
cosmological density fluctuations generated from the primordial
fluctuation of the curvaton.  Effects of the primordial fluctuation of
the inflaton will be discussed in Section \ref{sec:implications}.

Although we are interested in the case where the curvaton and the
Affleck-Dine field both exist, we first consider the adiabatic mode
(in the $\phi$D epoch and after) neglecting the effects of the
Affleck-Dine field.  Indeed, behaviors of the adiabatic mode can be
understood without taking account of the Affleck-Dine field as far as
the energy density of the Affleck-Dine field is always sub-dominant.

In this paper, we use the Newtonian gauge where the line element is
described as\footnote
{We use the convention and notation of \cite{Hu:1995em} unless
otherwise mentioned.}
\begin{eqnarray}
    ds^2 = - (1+2\Psi) dt^2 + (a/a_0)^2 (1+2\Phi) d{\bf x}^2.
\end{eqnarray}
Here, $a_0$ is some constant and ${\bf x}$ is the comoving coordinate.
In the RD1 and $\phi$D epochs, it is expected that the temperature of
the universe is so high that the momentum-exchange of the relativistic
particles are efficient enough.  In this case, anisotropic stress
vanishes and $\Phi =-\Psi$.  With this relation, relevant part of the
equations to be solved are, for the Fourier mode with the (comoving)
momentum $k$,
\begin{eqnarray}
    && 
    \dot{\Psi} + H \Psi = \frac{1}{2M_*^2}
    \left(
        \frac{3}{4} \frac{a}{a_0} 
        \rho_{\gamma_\chi} \tilde{V}_{\gamma_\chi}
        + \dot{\phi}\delta\phi \right),
    \label{Psidot}
    \\ &&
    \dot{\delta}_{\gamma_\chi} = 4 \dot{\Psi}
    - \frac{4}{3} \frac{a_0}{a} k^2 \tilde{V}_{\gamma_\chi},
    \label{drdot}
    \\ &&
    \dot{\tilde{V}}_{\gamma_\chi} = \frac{a_0}{a} 
    \left( \frac{1}{4} \delta_{\gamma_\chi} + \Psi \right),
    \\ &&
    \ddot{\delta\phi} + 3 H \dot{\delta\phi} 
    + \left[ \left( \frac{a_0}{a} \right)^2 k^2 + m_\phi^2 \right] 
    \delta\phi = 4 \dot{\phi} \dot{\Psi} - 2 m_\phi^2 \phi \Psi,
    \label{dphidot}
\end{eqnarray}
where
$\delta_{\gamma_\chi}\equiv\delta\rho_{\gamma_\chi}/\rho_{\gamma_\chi}$
with $\delta\rho_{\gamma_\chi}$ being the fluctuation of the energy
density of $\gamma_\chi$, $\tilde{V}_{\gamma_\chi}\equiv
V_{\gamma_\chi}/k$ with $V_{\gamma_\chi}$ being the velocity of
$\gamma_\chi$, and $\delta\phi$ is the fluctuation of the curvaton
amplitude.

In order to solve the equations, we have to specify the initial
conditions for the fluctuations (as well as those for
$\rho_{\gamma_\chi}$ and $\phi$).  In order to study the effects of
the primordial fluctuation of the curvaton, we concentrate on the mode
for which, in the deep RD1 epoch, $\phi$ has non-vanishing fluctuation
as given in Eq.\ (\ref{dphi_init}) while $\delta\rho_{\gamma_\chi}$
vanishes.

Importantly, we are interested in the modes related to currently
observed cosmological density fluctuations.  Those modes reenter the
horizon at the time close to the (usual) radiation-matter equality and
hence they are superhorizon modes in the RD1 and $\phi$D epochs.
Thus, in discussing their behaviors in the RD1 and $\phi$D epochs, all
the terms proportional to $k^2$ are irrelevant in the evolution
equations.  Then, solving Eqs.\ (\ref{Psidot}) $-$ (\ref{dphidot}),
fluctuations in the deep RD1 epoch when $H\gg m_\phi$ are given by
\begin{eqnarray}
    \Psi_{\rm RD1}^{(\delta\phi)} &\simeq& -\frac{1}{14} 
    \frac{\delta\rho_{\phi,{\rm init}}}{\rho_{\gamma_\chi}},
    \label{Psi_RD1}
    \\
    \delta_{\gamma_\chi,{\rm RD1}}^{(\delta\phi)} &\simeq& 
    -\frac{2}{7} 
    \frac{\delta\rho_{\phi,{\rm init}}}{\rho_{\gamma_\chi}},
    \\
    \tilde{V}_{\gamma_\chi, {\rm RD1}}^{(\delta\phi)} &\simeq& 
    -\frac{\sqrt{2H_0t}}{35} 
    \frac{\delta\rho_{\phi,{\rm init}}}{\rho_{\gamma_\chi}},
    \\
    \delta\phi_{\rm RD1}^{(\delta\phi)} &\simeq&
    \left( 1 - \frac{1}{5} m_\phi^2 t^2 \right)
    \delta\phi_{\rm init},
    \label{dphi_RD1}
\end{eqnarray}
where the subscript ``RD1'' is for variables in deep RD1 epoch, and
the superscript ``$(\delta\phi)$'' is for fluctuations generated from
the primordial fluctuation of the curvaton.  Here, $H_0$ is the
expansion rate at $a=a_0$ (which is, of course, taken at deep RD1
epoch), and
\begin{eqnarray}
    \delta\rho_{\phi,{\rm init}}\equiv 
    m_\phi^2 \phi_{\rm init}\delta\phi_{\rm init}.
\end{eqnarray}
Notice that the relations given in Eqs.\ (\ref{Psi_RD1}) $-$
(\ref{dphi_RD1}) provide initial conditions for our numerical
calculations.

Evolutions of the fluctuations can be studied by solving Eqs.\ 
(\ref{Psidot}) $-$ (\ref{dphidot}).  Importantly, in the curvaton
scenario, fluctuation in the curvaton sector is converted to the
adiabatic density fluctuations after the $\phi$D epoch.  Evolutions of
the superhorizon fluctuations are the same as those with the baryonic
isocurvature fluctuations \cite{Mollerach:hu}; after the $\phi$D
epoch, $\Psi^{(\delta\phi)}$ becomes of the order of
$S_{\phi\chi}^{(\delta\phi)}$, where
\begin{eqnarray}
    S_{\phi\chi}^{(\delta\phi)} \equiv 
    2 \frac{\delta\phi_{\rm init}}{\phi_{\rm init}}.
\end{eqnarray}
If the initial amplitude of the curvaton is much smaller than $M_*$,
$\phi$ starts to oscillate when the universe is dominated by
radiation.  In this case, $S_{\phi\chi}^{(\delta\phi)}$ becomes the
entropy fluctuation between components generated from the decay
products of $\phi$ and those generated from $\chi$ (like
$\gamma_\chi$).  Since the curvaton can be identified as a
non-relativistic component once it starts to oscillate, situation is
the same as the models with isocurvature fluctuation in
non-relativistic component.  Then we obtain
\begin{eqnarray}
    \Psi_{\rm RD2}^{(\delta\phi)} = 
    \frac{10}{9} \Psi_{\rm \phi D}^{(\delta\phi)} = 
    - \frac{4}{9} \frac{\delta\phi_{\rm init}}{\phi_{\rm init}},
    \label{Psi(RD2)}
\end{eqnarray}
where the subscripts ``$\phi$D'' and ``RD2'' are quantities in the
$\phi$D and RD2 epochs, respectively.

\subsection{Entropy fluctuation}

Now, we discuss the entropy fluctuation in the baryonic sector, which
is defined as
\begin{eqnarray}
    S_{b\gamma} = \frac{\delta(n_b/n_\gamma)}{(n_b/n_\gamma)}.
    \label{S_bg}
\end{eqnarray}
Although the precise calculation of the entropy fluctuation will be
done in the next section with numerical calculations, it is
instructive to discuss the basic behavior of the entropy fluctuation.
Indeed, in some case, baryonic entropy fluctuation associated with the
Affleck-Dine field can be analytically estimated.

Here, we consider the case where the energy density of the
Affleck-Dine field is sub-dominant in the RD1 and $\phi$D epochs.  In
such a case, Eqs.\ (\ref{eta_1}) and (\ref{eta_2}) are applicable,
from which we can estimate the entropy fluctuation generated in our
scenario.  As can be seen in Eq.\ (\ref{eta_1}), if $\phi_{\rm
init}\ll\phi_{\rm crit}$, $n_b/n_\gamma$ depends on $\phi_{\rm init}$.
Thus, if the curvaton amplitude has primordial fluctuation, it also
generates the entropy fluctuation between the baryon and the photon.
Using Eq.\ (\ref{S_bg}),
\begin{eqnarray}
    S_{b\gamma}^{(\delta\phi)} \simeq 
    -2 \frac{\delta\phi_{\rm init}}{\phi_{\rm init}}
    {\rm ~~for~~} \phi_{\rm init} \ll \phi_{\rm crit}.
\end{eqnarray}
On the contrary, when $\phi_{\rm init}\gg\phi_{\rm crit}$,
$n_b/n_\gamma$ is independent of $\phi_{\rm init}$, and we obtain
\begin{eqnarray}
    S_{b\gamma}^{(\delta\phi)} \simeq 0
    {\rm ~~for~~} \phi_{\rm init} \gg \phi_{\rm crit}.
\end{eqnarray}

Importantly, shape of the CMB angular power spectrum depends on the
relative size between the adiabatic and entropy fluctuations.  Thus,
for our following discussions, it is convenient to define
\begin{eqnarray}
    \kappa_b \equiv 
    \frac{S_{b\gamma}^{(\delta\phi)}}{\Psi_{\rm RD2}^{(\delta\phi)}}.
\end{eqnarray}
If the curvaton starts to oscillate in the deep RD1 epoch, which is
realized when $\phi_{\rm init}\ll M_*$, metric perturbation is also
calculated as given in Eq.\ (\ref{Psi(RD2)}) and 
obtain
\begin{eqnarray}
    \kappa_b \simeq
    \left\{ \begin{array}{ll}
            9/2 & ~~~:~~~ \phi_{\rm init} \ll \phi_{\rm crit} \\
            0   & ~~~:~~~ \phi_{\rm init} \gg \phi_{\rm crit}
        \end{array} \right. .
\end{eqnarray}

As mentioned before, the CMB angular power spectrum obtained by the
WMAP is highly consistent with the prediction of the purely adiabatic
primordial density fluctuations.  Thus, if $|\kappa_b|$ is too large,
resultant CMB angular power spectrum becomes inconsistent with the
observations.  Indeed, size of the correlated isocurvature fluctuation
is constrained to be \cite{HamKawMorTak}
\begin{eqnarray}
    |\kappa_b|\lesssim 0.5.
    \label{k<0.5}
\end{eqnarray}
With this constraint, it is obvious that the case with $\phi_{\rm
init}\ll\phi_{\rm crit}$ is excluded.  Consequently, if the
Affleck-Dine mechanism is implemented in the curvaton scenario, one of
the following two conditions should be satisfied: (i) $\phi_{\rm
init}\gtrsim M_*$, or (ii) $m_\psi\ll m_\phi$.  In order to be more
quantitative, in the next section, we use numerical method to
calculate $\kappa_b$ (and other quantities) and derive the
constraints.

Notice that, in the extreme case where the Affleck-Dine field
eventually dominates the universe after the decay of the curvaton, all
the components in the universe are generated from the Affleck-Dine
field.  In this case, entropy fluctuation vanishes and the cosmic
density fluctuations are purely adiabatic.  (See Eq.\ (\ref{eta_3}).)
Then, $\kappa_b=0$ and the resultant cosmic density fluctuations
become consistent with the observations.  Thus, hereafter, we
concentrate on the cases where the present CMB radiation (as well as
other components in the universe) are generated from the decay product
of $\phi$.\footnote
{In our discussion, for simplicity, it is assumed that the energy
density of the Affleck-Dine field is always sub-dominant.  Our results
are, however applicable to the case where the energy density of the
Affleck-Dine field becomes comparable to that of the curvaton at some
epoch as far as the present CMB radiation (as well as other components
in the universe) are generated from the decay product of $\phi$.}

Before closing this section, we comment on the effects of the
primordial fluctuation of the Affleck-Dine field.  If the effective
mass of the Affleck-Dine field is much smaller than the expansion rate
during the inflation, it is expected that the Affleck-Dine field also
acquires the primordial fluctuation $\delta\psi_{\rm init}$.  Since the
resultant baryon-to-photon ratio depends on the initial amplitude of
$\psi$, such a primordial fluctuation results in extra entropy
fluctuation in the baryonic sector.  Assuming no correlation between
the primordial fluctuations of the curvaton and the Affleck-Dine
field, $\delta\psi_{\rm init}$ generates uncorrelated baryonic entropy
fluctuation.  Using Eq.\ (\ref{eta_1}) with $\phi_{\rm init}\sim M_*$,
which is required in order to suppress $\kappa_b$ when $m_\phi\sim
m_\psi$ as will be seen in the next section, we obtain
\begin{eqnarray}
    S_{b\gamma}^{(\delta\psi)} \sim 
    \frac{\delta\psi_{\rm init}}{\psi_{\rm init}},
\end{eqnarray}
where $S_{b\gamma}^{(\delta\psi)}$ is the baryonic entropy fluctuation
associated with the primordial fluctuation of the Affleck-Dine field.
Defining 
\begin{eqnarray}
    \kappa_b^{\rm (uncorr)} \equiv 
    \frac{S_{b\gamma}^{(\delta\psi)}}{\Psi_{\rm RD2}^{(\delta\phi)}},
\end{eqnarray}
we obtain $\kappa_b^{\rm (uncorr)}\sim \phi_{\rm init}/\psi_{\rm
init}$.  Uncorrelated entropy fluctuation is constrained by the WMAP
results and $\kappa_b^{\rm (uncorr)}\lesssim 3$ \cite{HamKawMorTak}.
This constraint can be evaded if the initial amplitude of the
Affleck-Dine field is not much smaller than that of the curvaton.  Of
course, if $\psi$ acquires effective mass as large as the expansion
rate during the inflation, $\delta\psi_{\rm init}$ vanishes and there
is no (uncorrelated) baryonic entropy fluctuation.  Hereafter, we
consider the case where the effects of the uncorrelated baryonic
entropy fluctuation becomes not important, and concentrate on the
effects of correlated baryonic entropy fluctuation.

\section{Numerical Results}
\label{sec:numerical}
\setcounter{equation}{0}

\subsection{Metric perturbation}

So far, we have seen that the case with $\phi_{\rm init}\ll\phi_{\rm
crit}$ is inconsistent with the observations.  One possibility of
realizing $\phi_{\rm init}\gtrsim\phi_{\rm crit}$ when $m_\psi\sim
m_\phi$ is to adopt $\phi_{\rm init}\gtrsim M_*$.  In this case, (a
short period of) inflation may occur due to the energy density of the
curvaton field and Eq.\ (\ref{Psi(RD2)}) becomes unapplicable.  In
addition, if $\phi_{\rm init}\sim\phi_{\rm crit}$, analytic estimation
of the precise value of $\kappa_b$ is difficult.  Thus, in this
section, we use numerical method to calculate $\Psi$ and $S_{b\gamma}$
associated with the primordial fluctuation of the curvaton.

The first step is to calculate the metric perturbation generated from
$\delta\phi_{\rm init}$.  We use the initial conditions given in Eqs.\ 
(\ref{Psi_RD1}) $-$ (\ref{dphi_RD1}) in the deep RD1 epoch, and
numerically solve Eqs.\ (\ref{Psidot}) $-$ (\ref{dphidot}) as well as
the equations for the zero-mode from the RD1 epoch to $\phi$D epoch.
We checked that $\Psi$ becomes constant in the $\phi$D epoch.  Metric
perturbation in the RD2 epoch is evaluated with the relation
$\Psi^{(\delta\phi)}_{\rm RD2}=\frac{10}{9}\Psi^{(\delta\phi)}_{\rm
\phi D}$.

\begin{figure}
    \centerline{\epsfxsize=0.5\textwidth\epsfbox{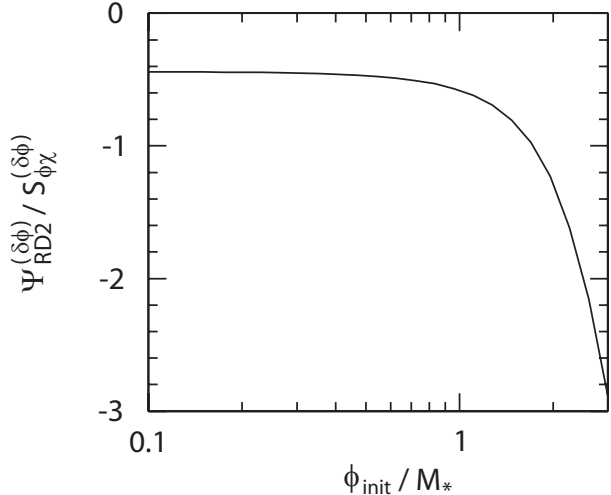}}
    \caption{$\Psi^{(\delta\phi)}_{\rm RD2}$ normalized by
    $S_{\phi\chi}^{(\delta\phi)}\equiv 2\delta\phi_{\rm
    init}/\phi_{\rm init}$ as a function of $\phi_{\rm init}$.}
    \label{fig:psird}
\end{figure}

In Fig.\ \ref{fig:psird}, we plot $\Psi^{(\delta\phi)}_{\rm RD2}$
normalized by $S_{\phi\chi}^{(\delta\phi)}\equiv 2\delta\phi_{\rm
init}/\phi_{\rm init}$ as a function of $\phi_{\rm init}$.  We have
checked that the ratio $\Psi^{(\delta\phi)}_{\rm
RD2}/S_{\phi\chi}^{(\delta\phi)}$ is independent of $m_\phi$.  As we
have discussed in the previous section, the ratio
$\Psi^{(\delta\phi)}_{\rm RD2}/S_{\phi\chi}^{(\delta\phi)}$ becomes
$-\frac{2}{9}$ when $\phi_{\rm init}\ll M_*$.  (See Eq.\ 
(\ref{Psi(RD2)}).)  In this case, the curvaton becomes (almost)
non-relativistic matter in the deep RD1 epoch.  Once $\phi_{\rm init}$
becomes close to $M_*$, on the contrary, $H\sim m_\phi$ is realized
when the energy density of $\phi$ becomes comparable to that of
radiation.  Then, $\Psi^{(\delta\phi)}_{\rm RD2}$ deviates from the
result given in Eq.  (\ref{Psi(RD2)}) and, as seen in Fig.\ 
\ref{fig:psird}, $|\Psi^{(\delta\phi)}_{\rm
RD2}/S_{\phi\chi}^{(\delta\phi)}|$ becomes larger than $\frac{2}{9}$.

\subsection{Baryonic entropy fluctuation}

Now, we consider the baryonic entropy fluctuation and discuss
constraints on the curvaton scenario with the Affleck-Dine
baryogenesis.  For this purpose, we numerically solve Eqs.\
(\ref{Psidot}) $-$ (\ref{dphidot}) simultaneously with the evolution
equation of the Affleck-Dine field $\psi$.  In our analysis, entropy
fluctuation is evaluated by taking the derivative numerically;
denoting the resultant baryon and curvaton number densities with the
initial amplitude $\phi_{\rm init}$ as $n_b(\phi_{\rm init})$ and
$n_\phi(\phi_{\rm init})\equiv\rho_\phi(\phi_{\rm init})/m_\phi$,
respectively, the entropy fluctuation between the baryon and the
radiation is calculated as
\begin{eqnarray}
    S_{b\gamma}^{(\delta\phi)} = 
    \frac{
    n_b(\phi_{\rm init}+\delta\phi_{\rm init})/
    n_\phi(\phi_{\rm init}+\delta\phi_{\rm init}) - 
    n_b(\phi_{\rm init})/n_\phi(\phi_{\rm init})}
    {n_b(\phi_{\rm init})/n_\phi(\phi_{\rm init})}.
\end{eqnarray}

\begin{figure}
    \centerline{\epsfxsize=0.5\textwidth\epsfbox{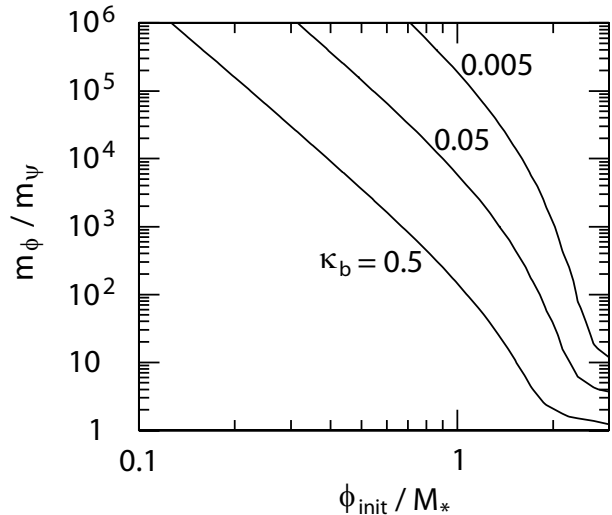}}
    \caption{Contours of constant $\kappa_b$ on
    $m_\phi/m_\psi$ vs.\ $\phi_{\rm init}$ plane.  The contours
    correspond to $\kappa_b=0.5$, $0.05$, and $0.005$
    from below.  We take $m_\psi=100\ {\rm GeV}$.  The parameter
    region below the contour for $\kappa_b=0.5$ is
    inconsistent with the WMAP result.}
    \label{fig:kappab}
\end{figure}

With the entropy fluctuation in the baryonic sector, we can calculate
the $\kappa_b$ parameter.  The result is shown in Fig.\ 
\ref{fig:kappab} on the $\phi_{\rm init}$ vs.\ $m_\phi$ plane.  The
metric perturbation $\Psi_{\rm RD2}^{(\delta\phi)}$ is determined by
$\phi_{\rm init}$ (with $S_{\phi\chi}^{(\delta\phi)}$ being fixed),
and has relatively mild dependence on $\phi_{\rm init}$.  Thus, Fig.\ 
\ref{fig:kappab} primarily shows the behavior of baryonic entropy
fluctuation in this model.  With this in mind, the behavior of
$\kappa_b$ can be easily understood.  The important feature is that
$S_{b\gamma}^{(\delta\phi)}\rightarrow 0$ as $\phi_{\rm init}$ or
$m_\phi$ becomes large enough.  This is because, as explained in the
previous section, baryonic entropy fluctuation is suppressed if the
Affleck-Dine field starts to oscillate after the $\phi$D epoch is
realized.  In other words, if the expansion rate at the time of the
radiation-curvaton equality is larger than $m_\psi$, resultant entropy
fluctuations become adiabatic.  Since $H_{\rm eq}\propto
m_\phi\phi_{\rm init}^4$, $m_\phi$ or $\phi_{\rm init}^4$ is required
to be large enough for suppressing baryonic entropy fluctuations.  As
a result, $\kappa_b$ becomes close to zero when the combination
$m_\phi$ or $\phi_{\rm init}$ becomes large.

With the constraint on the $\kappa_b$ parameter from the WMAP
(\ref{k<0.5}), we obtain stringent constraint on the initial amplitude
and the mass of the curvaton field.  In particular, for a fixed value
of the initial amplitude $\phi_{\rm init}$, we obtain an lower bound
on $m_\phi$.   Implications of such a constraint will be discussed in
the next section.

\section{Implications}
\label{sec:implications}
\setcounter{equation}{0}

In the previous sections, we have studied general constraints without
specifying the detailed properties of the curvaron field.
Consequently, we have seen that the expansion rate at the time of the
radiation-curvaton equality should be larger than the mass of the
Affleck-Dine field; otherwise, the Affleck-Dine field starts to move
in the RD1 epoch and too large baryonic entropy fluctuaion is
generated.  This fact has serious implications to some of the cases.

In particular, it is important to consider the case where the mass of
the curvaton (as well as that of the Affleck-Dine field) is from the
effect of the supersymmetry breaking.  In this case, mass of the
curvaton is expected to be close to that of the Affleck-Dine field and
we obtain serious constraint on the initial amplitude of the curvaton.

One of the important cases is that the cosmological modulus field
plays the role of the curvaton.  The moduli fields in the superstring
theory acquire masses from the effects of the supersymmetry breaking
and hence their masses are expected to be of the order of the
gravitino mass.  In addition, their interactions are expected to be
suppressed by inverse powers of the gravitational scale $M_*$ and
hence, once their amplitudes become smaller than $M_*$, their
potentials are well approximated by the parabolic ones.  Thus, if one
of such moduli fields has non-vanishing amplitude as well as the
primordial fluctuaion, it may play the role of
the curvaton.

If the mass of the modulus field (which is close to the gravitino
mass) is of $O(100\ {\rm GeV})$ as in the case of the
supergravity-induced supersymmetry breaking, its lifetime becomes much
longer than $1\ {\rm sec}$.  With such a long lifetime, decay of the
curvaton occurs after the big-bang nucleosynthesis (BBN) and it spoils
the success of the BBN \cite{bbn}.  If the modulus mass becomes larger
than $\sim (10-100)\ {\rm TeV}$, on the contrary, the cosmological
modulus field may decay before the BBN and it may be a viable
candidate of the curvaton.  In the scenario of the anomaly-mediated
supersymmetry breaking \cite{amsb}, for example, this may be the case.
In addition, in the anomaly-mediated models, Wino may become the
lightest superparticle; importantly, non-thermally produced Winos from
the decay of the cosmological modulus field (i.e., the curvaton) can
become the cold dark matter \cite{Moroi:1999zb}.  In such a model,
baryon asymmetry of the universe should be generated with very low
reheating temperature and with very large dilution factor; as a
candidate of the scenario of baryogenesis in such a situation, the
Affleck-Dine mechanism has been known to be promising
\cite{Moroi:1994rs}.

Although a sizable hierarchy is possible between $m_\phi$ and $m_\psi$
in the anomaly-mediated scenarios, we still obtain remarkable
constraint on $\phi_{\rm init}$.  Even for the case with relatively
large hierarchy between $m_\phi$ and $m_\psi$ as $m_\phi/m_\psi=10^2$
($10^3$), for example, $\phi_{\rm init}$ should be larger than
$1.1M_*$ ($0.7M_*$); otherwise, too large baryonic entropy fluctuation
is generated as seen in Fig.\ \ref{fig:kappab}.  For the case of the
anomaly-mediated supersymmetry breaking, the mass of the Affleck-Dine
field, which is given by the masses of the MSSM particles, is
suppressed by the loop-induced factor of order $\sim 10^{-2}-10^{-3}$
compared to the gravitino mass.  Even in that case, $\phi_{\rm
init}\sim M_*$ is required in order to suppress the baryonic entropy
fluctuations.  

In fact, if the curvaton field has an initial amplitude as large as
$\sim M_*$, it becomes non-trivial if the inflaton contribution to the
cosmic density fluctuations is negligible.  Importance of the
inflaton contribution can be seen by comparing the metric
perturbations generated from the fluctuations of the inflaton and the
curvaton.  The curvaton contribution is given in Eq.\ (\ref{Psi(RD2)})
while the inflaton contribution is given by
\begin{eqnarray}
    \Psi_{\rm RD2}^{\rm (inf)} = 
    \frac{2}{3}
    \left[ 
        \frac{3H_{\rm inf}^2}{V'_{\rm inf}} \times 
        \frac{H_{\rm inf}}{2\pi} 
    \right]_{k=aH},
\end{eqnarray}
where $V_{\rm inf}$ is the inflaton potential, $V_{\rm
inf}'\equiv\partial V_{\rm inf}/\partial\chi$, and the superscript
``(inf)'' is for variables generated from the primordial fluctuation
of the inflaton field.  If $|\Psi_{\rm
RD2}^{(\delta\phi)}|\gg|\Psi_{\rm RD2}^{\rm (inf)}|$, the curvaton
contribution dominates.  The relative size depends on the model of the
inflation.

For example, for the case of the chaotic inflation with inflaton
potential of the form $V_{\rm inf}\propto\chi^{q}$, we obtain
\begin{eqnarray}
    \frac{\Psi_{\rm RD2}^{\rm (inf)}}{\Psi_{\rm RD2}^{(\delta\phi)}} 
    =
    \left[ \frac{3}{2q} 
        \frac{\phi_{\rm init}\chi}{M_*^2} \right]_{k=aH_{\rm inf}},
\end{eqnarray}
where we have used Eq.\ (\ref{Psi(RD2)}) as an approximation.  (In
fact, as shown in Fig.\ \ref{fig:psird}, $|\Psi_{\rm
RD2}^{(\delta\phi)}|$ becomes larger than the value given in Eq.\ 
(\ref{Psi(RD2)}) if the initial amplitude of the curvaton becomes
comparable to $M_*$.  Then, the ratio becomes smaller.)  Using the
fact that the inflaton amplitude is $\sim 15M_*$ when the COBE scale
exits the horizon, $\Psi_{\rm RD2}^{\rm (inf)}$ becomes comparable to
or larger than $\Psi_{\rm RD2}^{(\delta\phi)}$ for $q=2-6$ even if
$\phi_{\rm init}\sim M_*$. Thus, sizable fraction of the cosmological
density fluctuation originates from the inflaton fluctuation; with the
chaotic inflation, the inflaton contribution is non-negligible unless
there is a large hierarchy between the masses of the curvaton and the
Affleck-Dine field.  In other words, if we consider the case where
$m_\phi$ is close to $m_\psi$, $\phi_{\rm init}\sim M_*$ is required
and the cosmic density fluctuations may not be dominated by the
curvaton contribution.

Assuming no correlation between the inflaton and curvaton fields, the
CMB angular power spectrum has the form
\begin{eqnarray}
    C_l = C_l^{(\delta\phi)} + C_l^{\rm (inf)},
\end{eqnarray}
where $C_l^{(\delta\phi)}$ and $C_l^{\rm (inf)}$ are the CMB angular
power spectra for the cases where the primordial fluctuations of the
curvaton and inflaton dominates, respectively.  As we mentioned,
$C_l^{\rm (inf)}$ may become comparable to (or even larger than)
$C_l^{(\delta\phi)}$.  Since the density fluctuations related to the
primordial fluctuation of the inflaton field are adiabatic,
constraints on $\phi_{\rm init}$ (for fixed value of $m_\phi/m_\psi$)
can be relaxed as far as the scale dependence of the primordial
density fluctuation is negligible.  On the contrary, however, if
$\Psi_{\rm RD2}^{\rm (inf)}$ becomes comparable or larger than
$\Psi_{\rm RD2}^{(\delta\phi)}$, significant amount of the cosmic
density fluctuations are generated from the primordial fluctuation of
the inflaton.  In this case, of course, it becomes difficult to relax
the observational constraints on the inflaton potential with the
curvaton, which has been one of the important motivation of the
curvaton scenario.

Of course, the chaotic model is not the only possibility of the
inflation and, in other case, inflaton contribution to the total
cosmic density fluctuations may become minor even though $\phi_{\rm
init}\sim M_*$.  Model of the inflation with large $H_{\rm inf}$ and
small $\Psi^{\rm (inf)}$ is considered, for example, in
\cite{Pilo:2004mg}.

\section{Summary}
\label{sec:summary}
\setcounter{equation}{0}

In this paper, we have discussed the curvaton scenario with
Affleck-Dine baryogenesis.  Even with a large entropy production due
to the decay of the curvaton field, Affleck-Dine mechanism may be able
to generate large enough baryon asymmetry and hence, for some of the
curvaton scenarios, Affleck-Dine mechanism is a prominent candidate of
the scenario of baryogenesis.  We have seen, however, that baryonic
entropy fluctuation is induced if the Affleck-Dine field starts to
move in the RD1 epoch.  Since large baryonic entropy fluctuation is
inconsistent with the observation (in particular, with the results of
the WMAP), this provides constraints on the scenario; as we have seen,
in order to evade the constraint, mass of the curvaton or the initial
amplitude of the curvaton should be large enough (for fixed value of
$m_\psi$).

We have also discussed implications of such constraints on some
scenario of the curvaton.  One of the cases where the Affleck-Dine
scenario is preferred is that the cosmological modulus field plays the
role of the curvaton; in such a case, baryon asymmetry should be
generated with large amount of entropy production and with very low
reheating temperature.  Assuming the anomaly-mediated supersymmetry
breaking, the ratio $m_\phi/m_\psi$ is expected to be $10^2-10^3$.
With such a ratio, we have seen that the initial amplitude of the
curvaton is constrained to be larger than $\sim M_*$ in order not to
generate too large baryonic entropy fluctuation.

{\sl Acknowledgments:} We acknowledge the use of CMBFAST
\cite{cmbfast} package for our numerical calculations.  The work of
T.M. is supported by the Grant-in-Aid for Scientific Research from the
Ministry of Education, Science, Sports, and Culture of Japan, No.\
15540247.

\end{document}